\documentclass[12pt,epsf,fleqn]{article}
\usepackage{epsfig}
\usepackage[dvips]{color}
\begin{document}
\title{Phase transition in a supersymmetric axion model}
\author{S. W. Ham$^{(1)}$ and S. K. Oh$^{(1,2)}$
\\
\\
{\it $^{(1)}$ Center for High Energy Physics, Kyungpook National University}\\
{\it Daegu 702-701, Korea} \\
{\it $^{(2)}$ Department of Physics, Konkuk University, Seoul 143-701, Korea}
\\
\\
}
\date{}
\maketitle
\begin{abstract}
In a supersymmetric axion model where the scale for both supersymmetry breaking and Peccei-Quinn symmetry
breaking is around $10^{11}$ GeV, we find that there is a reasonable parameter space for a strongly first
order phase transition at the scale.
\end{abstract}

\section{Introduction}

The existence of phase transition associated with spontaneous symmetry breaking may appear
during the evolution of the universe.
Such a phase transition may influence to the large-scale structure of the universe [1].
Both the order of the phase transition and its strength are basic ingredients
for a quantitative discussion of the transition at some energy scale.
In general, a symmetry-breaking phase transition can be first or second order one.
In a first order phase transition, there is a potential barrier between the symmetric phase state and
the broken phase state at the critical temperature.
As spontaneous symmetry is restored at a high temperature, the relevant particle becomes massless.

Recently, Barger and his colleagues have studied axion models with high-scale supersymmetry (SUSY)
breaking [2].
An axion associated with spontaneously broken Peccei-Quinn (PQ) symmetry gives a natural solution
to the strong CP problem [3].
The scale of decay constant of the Weinberg-Wilczek axion [4] is assumed to be the electroweak scale,
which has been excluded by negative result of experimental searches.
On the other hand, in the invisible axion models such as the Kim-Shifman-Vainshtein-Zakharov (KSVZ)
model [5] or the Dine-Fischler-Srednicki-Zhitnitskii (DFSZ) model [6],
the scale of decay constant of the invisible axion is assumed to be very high and
thus the invisible axion may exist within experimental constraints.
Both the mass and the coupling strength of the invisible axion are very small
since they are inversely proportional to the decay constant.

In the supersymmetric KSVZ axion model studied in Ref. [2], the intermediate scale of supersymmetry
breaking ($\sim 10^{11}$ GeV) is directly related to the PQ symmetry breaking scale.
The Higgs sector of this model consists of two Higgs doublets and one complex Higgs singlet.
At very high energy scale, there are one heavy scalar Higgs boson, two pairs of SM vector-like particles
in this model.
The mass of the heavy scalar Higgs boson of the supersymmetric KSVZ axion model is about $10^{11}$ GeV,
the SUSY breaking scale.
At the electroweak scale,  the supersymmetric KSVZ axion model has one light scalar Higgs boson.
The range of the light scalar Higgs boson mass is 130 to 160 GeV, after the squark, sleptons, gauginos,
and higgsinos are decoupled at the SUSY breaking scale.
Also, in this model there are supersymmetric particles, as well as the SM fermions the SM gauge bosons,
and two pairs of SM vector-like particles.

In this paper, we study the possibility of phase transition at the SUSY or PQ symmetry breaking scale
in the supersymmetric KSVZ axion model.
We find that there is a reasonable parameter space for a strongly first order phase transition
at the energy scale of the SUSY or PQ symmetry breaking in the KSVZ axion model.

\section{The supersymmetric KSVZ model}

We first describe briefly the supersymmetric KSVZ model, following the notations of Ref. [2].
The left-handed quark doublets, the right-handed up-type quarks, the right-handed down-type quarks,
the left-handed lepton doublets, and the right-handed leptons are denoted as $Q_i$, $U_i^c$, $D_i^c$, $L_i$,
and $E_i^c$, respectively, where $i$ is the family index.
The two Higgs doublets and one complex Higgs singlet are denoted as $H_u$, $H_d$, and $S$, respectively.
In addition, in the supersymmetric KSVZ model, there are two pairs of
SM vector-like particles ($Q_X, {\bar Q}_X$) and ($D_X, {\bar D}_X$).
We would not consider the lowest higher-dimensional operator in the superpotential,
which is suppressed at the Planck scale.

The superpotential of the supersymmetric KSVZ model is given by
\begin{equation}
    W = y_{ij}^u H_u \epsilon U_i^c Q_j - y_{ij}^d H_d \epsilon D_i^c Q_j
        -  y_{ij}^w H_d \epsilon E_i^c L_j
        + y_{Q_X} S Q_X {\bar Q}_X
        + y_{D_X} S D_X {\bar D}_X ,
\end{equation}
where $y_{ij}^u$, $y_{ij}^d$, $y_{ij}^e$, $y_{Q_X}$, and $y_{D_X}$ are the Yukawa couplings
for the relevant particles, and $\epsilon \equiv i \sigma^2$, where $\sigma^2$ is the second Pauli matrix.
From the two Higgs doublets, two neutral scalar Higgs bosons emerge.
One of them becomes the SM-like scalar Higgs boson, which is fine-tuned to have a small mass
in the range between 130 and 160 GeV, whereas the mass of the other one is comparable
to the SUSY breaking scale ($10^{11}$ GeV).

In the supersymmetric KSVZ axion model, the SUSY and the PQ symmetry are broken
at an intermediate energy scale of about $10^{11}$ GeV.
We concentrate on the relevant effective potential for $S$ at this energy scale.
We denote the soft SUSY breaking masses as ${\tilde m}_{Q_X}$ for both $Q_X$ and ${\bar Q}_X$,
and  as ${\tilde m}_{D_X}$ for both $D_X$ and ${\bar D}_X$.
We assume that $y_{Q_X} = y_{D_X}$ and ${\tilde m}_{Q_X} = {\tilde m}_{D_X}$.
At zero temperature, the one-loop radiative corrections associated with $S$ come from
the SM vector-like particles.
Thus, the one-loop effective potential is given by [7, 8]
\begin{equation}
    V_1 (S, 0) =
    - {\tilde m}_S^2 S^2 + {C_+ \over 64 \pi^2} \int_0^{\Lambda^2} d k^2 k^2
    \left \{ \log \left (1 + {y_{Q_X}^2 S^2 \over k^2 + {\tilde m}_{Q_X}} \right )
    - \log \left (1 + {y_{Q_X}^2 S^2 \over k^2} \right )
    \right \} ,
\end{equation}
where $C_+$ is a constant  coming from a product of color, charge, and spin factors
and the relevant particle number, and ${\tilde m}_S$ is the soft SUSY breaking mass for $S$.
Note that both the vacuum expectation value (VEV) of the Higgs singlet and the soft SUSY breaking mass
are about $10^{11}$ GeV. Neglecting terms that vanish as $\Lambda \rightarrow \infty$,
we obtain the approximate formula for the one-loop effective potential as
\begin{eqnarray}
    V_1 (S, 0)
    & = &\mbox{}
    - {\tilde m}_S^2 S^2 - {C_+ y_{Q_X}^4 S^4 \over 64 \pi^2} \log (y_{Q_X}^2 S^2)  \cr
    & & \mbox{}
    + {C_+ \over 64 \pi^2} ({\tilde m}_{Q_X}^2  + y_{Q_X}^2 S^2)^2
    \log ({\tilde m}_{Q_X}^2 + y_{Q_X}^2 S^2) \cr
    & & \mbox{}
    - {C_+ \over 64 \pi^2} {\tilde m}_{Q_X}^2 y_{Q_X}^2 S^2 (2 \log (\Lambda^2) + 1)  ,
\end{eqnarray}
where the soft SUSY breaking mass ${\tilde m}_S$ can be eliminated by the minimization of the potential.
At the one-loop level, the the soft SUSY breaking mass is expressed as
\begin{equation}
{\tilde m}_S^2 =  {C_+ y_{Q_X}^4 f_a^2 \over 32 \pi^2}
     \log \left (1 + { {\tilde m}_{Q_X}^2 \over y_{Q_X}^2 f_a^2} \right )
    + {C_+ y_{Q_X}^2 {\tilde m}_{Q_X}^2 \over 32 \pi^2}
    \log \left ({y_{Q_X}^2 f_a^2 + {\tilde m}_{Q_X}^2 \over \Lambda^2} \right ) ,
\end{equation}
where the VEV of the Higgs singlet have $f_a = 10^{11}$ GeV.

Now, the finite temperature contribution to the one-loop potential is given by [9]
\begin{eqnarray}
    V_1 (S, T)
    & = &\mbox{}
    - {C_+ T^4 \over 2 \pi^2} \int_0^{\infty} dx \ x^2 \ \log
    \left [1+ \exp{\left ( - \sqrt {x^2 + {y_{Q_X}^2 S^2/T^2 }} \right )  } \right ] \cr
    &  &\mbox{}
    + {C_+ T^4 \over 2 \pi^2}  \int_0^{\infty} dx \ x^2 \ \log
    \left [1- \exp{\left ( - \sqrt {x^2 + {( {\tilde m}_{Q_X}^2+ y_{Q_X}^2 S^2)/T^2 }} \right )  }
    \right ] ,
\end{eqnarray}
where the temperature is about $10^{11}$ GeV since the energy scale of both the VEV of $S$
and ${\tilde m}_{Q_X}$ is around $10^{11}$ GeV.
The SUSY breaking scale should be larger than $10^{11}$ GeV in order to obtain a reasonable value
of the quartic coupling [2].
Note that both $V_1 (S, 0)$ and $V_1 (S, T)$ vanish as the soft SUSY breaking mass goes to zero.
The full one-loop effective potential at finite temperature can be expressed
as $V (S,T) = V_1 (S,0) + V_1 (S,T)$.

\setcounter{figure}{0}
\def\figurename{}{}%

\renewcommand\thefigure{Fig. 1}
\begin{figure}[t]
\begin{center}
\includegraphics[scale=0.9]{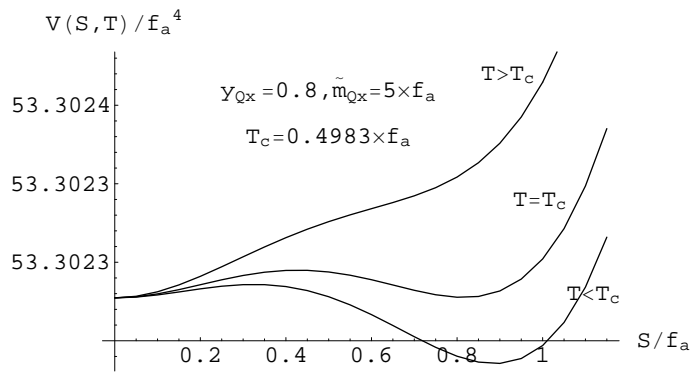}
\caption[plot]{The behavior of the one-loop effective potential at finite temperature
as a function of $S/f_a$, for $y_{Q_X} = 0.8$ and ${\tilde m}_{Q_X} = 5 f_a$
for three typical temperatures, $T < T_c$, $T = T_c$, and $T > T_c$.
As the temperature increases, the spontaneous symmetry breaking disappears, and vice versa.
At the critical temperature, $T_c = 0.4983 f_a$, the temperature-dependent one-loop effective potential
exhibit two degenerate vacua.}
\end{center}
\end{figure}

We calculate the behavior of the full one-loop effective potential at finite temperature $V(S,T)$
as a function of $S/f_a$, for $y_{Q_X} = 0.8$ and ${\tilde m}_{Q_X} = 5 f_a$, where $\lambda_S = 0.17$, for three different temperatures.
Our result is shown in Fig. 1.
One can see that, for high temperature, namely, $T > T_c$, the potential is symmetric with one true vacuum.
On the other hand, for low temperature, namely, $T < T_c$, the symmetry of the potential is
spontaneously broken where the true vacuum occurs not at $S/f_a = 0$ but at somewhere with nonzero $S/f_a$.
In between, there is the critical temperature, $T = T_c$,
where the temperature-dependent effective potential has two degenerate vacua.
Our calculation yields $T_c = 0.4983 f_a$.
The nonzero value of $S$ for the two degenerate vacua at the critical temperature is $S_c= 0.8 f_a$.
This in turn yields $S_c/ T_c \approx 1.6$.
The criteria of a strongly first order phase transition is given by $v_c/T_c \ge 1$, where $v_c$ is
the nonzero value of the scalar field for the two degenerate vacua.
In our case, $S_c$ has the meaning of $v_c$.
Thus, Fig. 1 tells us that there can be a strongly first order phase transition
in the supersymmetric KSVZ model, for the parameter values we set.

Now, we calculate $S_c/T_c$ by varying the soft SUSY breaking mass ${\tilde m}_{Q_X}$.
The result is displayed in Table I, for $y_{Q_X} = 0.8$.
Table I shows that as ${\tilde m}_{Q_X}$ increases strength of the phase transition gets weaker
until $S_c/T_c$ decreases down below 1.0 such that the phase transition becomes weakly first order
for ${\tilde m}_{Q_X} > 20 f_a$.
The quartic coupling for the Higgs singlet $S$ is given approximately by
\begin{equation}
    \lambda_S  = {9 y_{Q_X}^4 \over 8 \pi^2}
    \log \left (1 + {{\tilde m}_{Q_X}^2 \over y_{Q_X}^2 f_a^2} \right ) ,
\end{equation}
where $f_a$ is the VEV of $S$.
For $y_{Q_X} = 0.8$ and ${\tilde m}_{Q_X} = 5 f_a$, the quartic coupling $\lambda_S$ is 0.17.
We note that a reasonable value for $\lambda_S$ cannot be obtained if ${\tilde m}_{Q_X} < 0.1 f_a$.
We also note that the strength of the phase transition is enhanced
by the temperature-dependent one-loop effective potential through the contributions
from the neutral Higgs boson associated with the Higgs singlet.

\begin{table}[ht]
\caption{Strength of the first order phase transition for SUSY breaking scale
$0.01 \le {\tilde m}_{Q_X} /f_a \le 20$ for $y_{Q_X} = 0.8$. If $S_c/T_c > 1$,
the phase transition is regarded as strongly first order.}
\begin{center}
\begin{tabular}{c|c|c|c}
\hline
\hline
${\tilde m}_{Q_X}/f_a$ & $T_c/f_a$  & $S_c/f_a$ & $S_c/T_c$ \\
\hline
\hline
20 & 0.612 & 0.612 & 1  \\
\hline
10 & 0.5581 & 0.7 & 1.2  \\
\hline
5 & 0.4983 & 0.8 & 1.6  \\
\hline
1 & 0.312 & 1.1 & 3.5 \\
\hline
0.1 & 0.101 & 1.4 & 13.8 \\
\hline
0.01 & 0.032 & 1.4 & 43 \\
\hline
\hline
\end{tabular}
\end{center}
\end{table}

\section{Conclusions}

We examine for the possibility of strongly first order phase transition
in the supersymmetric KSVZ model proposed by Barger and colleagues, where a complex Higgs singlet
and two pairs of SM vector-like particles are introduced at a high scale associated with the SUSY
and PQ symmetry breakings.
We find that there is a parameter space in the model where a strongly first order phase transition
is possible at the SUSY breaking scale of the model, $10^{11}$ GeV.

\vskip 0.3 in

\noindent
{\large {\bf Acknowledgments}}
\vskip 0.2 in
\noindent
This research is supported through the Science Research Center Program
by the Korea Science and Engineering Foundation and the Ministry of Science and Technology.

\vskip 0.2 in



\end{document}